\documentclass[journal=jacsat,manuscript=article]{achemso}

\usepackage[version=3]{mhchem} 
\usepackage{tikz}
\newcommand*\circled[1]{\tikz[baseline=(char.base)]{
            \node[shape=circle,draw,inner sep=0.7pt] (char) {#1};}}

\author{L. Farcis}
\email{louis.farcis@cea.fr}
\author{B.M.S. Teixeira}%
\author{P. Talatchian}
\author{D. Salomoni}
\author{U. Ebels}
\author{S. Auffret}
\author{B. Dieny}
\affiliation[Spintec]{Univ. Grenoble Alpes, CEA, CNRS, Grenoble-INP, SPINTEC, 38000 Grenoble, France}%

\author{A. Mizrahi}
\author{J. Grollier}
\affiliation{Unité Mixte de Physique CNRS/Thales, Université Paris-Saclay, 91767 Palaiseau, France}%

\author{R.C. Sousa}
\author{L.D. Buda-Prejbeanu}
\affiliation[Spintec]{Univ. Grenoble Alpes, CEA, CNRS, Grenoble-INP, SPINTEC, 38000 Grenoble, France}%

\title[Spiking dynamics in dual free layer perpendicular magnetic tunnel junctions]{Spiking dynamics in dual free layer perpendicular magnetic tunnel junctions}

\begin{document}

\begin{abstract}

Spintronic devices have recently attracted a lot of attention in the field of unconventional computing due to their non-volatility for short and long term memory, non-linear fast response and relatively small footprint. Here we report how voltage driven magnetization dynamics of dual free layer perpendicular magnetic tunnel junctions enable to emulate spiking neurons in hardware. The output spiking rate was controlled by varying the dc bias voltage across the device. The field-free operation of this two-terminal device and its robustness against an externally applied magnetic field make it a suitable candidate to mimic neuron response in a dense Neural Network (NN). The small energy consumption of the device (4-16 pJ/spike) and its scalability are important benefits for embedded applications. This compact perpendicular magnetic tunnel junction structure could finally bring spiking neural networks (SNN) to sub-100nm size elements.

\end{abstract}

Neuromorphic computing is a promising way to drastically reduce the energy consumption of artificial intelligence, by building systems 
that are inspired by the structure and function of the brain to process information more efficiently in hardware \cite{markovic_physics_2020}. 
While biological neurons emit voltage spikes, most software artificial neural networks are static and rate-based, meaning that the outputs of neurons encode a rate of spikes rather than spike themselves. 
However, spiking neural networks have several important advantages for further reducing energy consumption. The unitary nature of spikes enables spiking neural networks with sparse activity, as well as asynchronous and event-based communication \cite{tavanaei_deep_2019}. Furthermore, electrical spikes can modify the conductance state of nanodevice-based artificial synapses, making it possible to implement local learning rules, a critical ingredient for on-chip learning \cite{serrano-gotarredona_stdp_2013, martin_eqspike_2021}.
Spiking neurons have been implemented using several substrates: CMOS circuits \cite{davies_loihi_2018,furber_spinnaker_2014, benjamin_neurogrid_2014}, photonic devices \cite{de_marinis_photonic_2019} and memristive devices \cite{nandakumar_experimental_2020}. 
However, spintronic spiking neurons are lacking. Spintronic-based neurons and synapses are a promising solution showing a significant reduction in energy consumption compared to CMOS-based circuits \cite{mizrahi_neural-like_2018}. Here we present an experimental demonstration of a spiking neuron implemented with a magnetic tunnel junction. Magnetic tunnel junctions combine the advantage of small size, high speed, low power consumption and CMOS compatibility. Our novel structure is based on a perpendicular magnetic tunnel junction (p-MTJ) with a dual-free layer configuration.
Using nanopillars from 70 to 150 nm of diameter, we demonstrate experimentally the windmill dynamics anticipated in simulation by Matsumoto \textit{et al.} \cite{matsumoto_chaos_2019}, without the need of a synthetic antiferromagnet (SAF) structure \cite{choi_current-induced_2016} and with a higher magnetoresistance signal than spin-valve structures \cite{thomas_spin_2017}.
We use electrical characterization to show the high tunability of the firing rate with respect to the bias voltage across the device. Furthermore, we develop a generalized compact model which successfully describes the evolution of the incubation time between spikes with respect to the applied voltage in the presence of an external applied field.

\begin{figure}
\includegraphics[width = \textwidth]{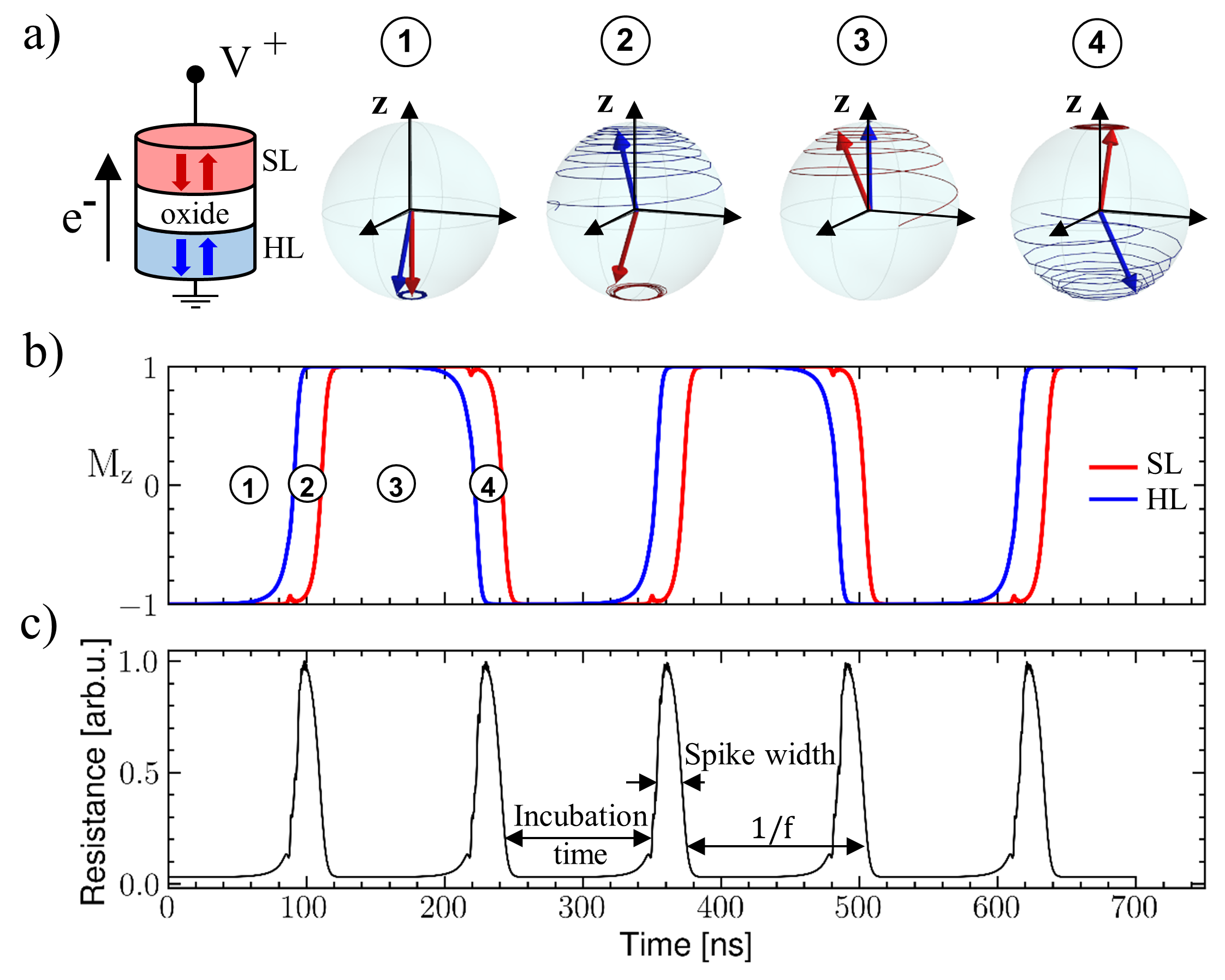}
\caption{\label{fig:1} a) Schematic view of dual-free layer p-MTJ and 3D macrospin configurations corresponding to several stages of the windmill dynamics. Simulated time traces of b) the  projection of the magnetization along z axis and c) the resulting normalized resistance under a bias voltage.}
\end{figure}

The structure of the spiking magnetic tunnel junction consists of two thin ferromagnetic layers with switchable magnetization oriented Out-Of-Plane (OOP) due to the perpendicular anisotropy induced at the interface with the Mg oxide spacer layer separating them \cite{dieny_perpendicular_2017, kishi_lower-current_2008, ikeda_perpendicular-anisotropy_2010} (see Fig.~\ref{fig:1}.a). 
Depending on the relative orientation of the magnetization of the two ferromagnetic layers, the junction resistance changes due to different tunneling probabilities of spin \textit{up} and spin \textit{down} electrons across the barrier, also called Tunnel Magneto Resistance effect \cite{julliere_tunneling_1975}. 
Here the two ferromagnetic layers have different anisotropy, leading to  different stability. They are therefore called hard layer (HL) and soft layer (SL).
Under sufficient applied voltage bias, dynamical states are created in which both hard layer (HL) and soft layer (SL) can switch by Spin-Transfer Torque (STT) \cite{slonczewski_current-driven_1996, berger_emission_1996}. 

To illustrate this dynamics, two coupled macrospins (SL and HL) have been simulated with the effective anisotropy of SL lower than the one of HL (see supplementary material 1). 
The switching sequence of Fig.~\ref{fig:1}.b can be explained as follows: from the parallel (P) alignment of SL and HL pointing down $\downarrow$, the reflected electrons (minority spin-up) at the interface between HL and the barrier forces the reversal of the HL from $\downarrow$ to $\uparrow$, step \circled{1} $\rightarrow$ \circled{2}.
Subsequently, the SL reverses thanks to the spin-torque coming from the majority spin-up electrons of the HL passing through the barrier. Since the effective anisotropy of SL is smaller than the HL one, the reversal of the SL \circled{2} $\rightarrow$ \circled{3} is faster. 
The two magnetizations retrieve a parallel alignment in the opposite up-direction  $\uparrow$ with respect to the starting point \circled{1}. Without applying any external magnetic field, the reversal process \circled{3}$ \rightarrow$ \circled{4} $\rightarrow$ \circled{1} is identical to \circled{1} $\rightarrow$ \circled{2} $\rightarrow$ \circled{3}. These magnetization switching sequences lead to a spiking behaviour of the electrical resistance. The characteristic lifetimes of the high and low resistance states are associated to the spike width and to the incubation time before spiking as illustrated in Fig.~\ref{fig:1}.c. In the following, the dynamics under a perpendicular applied field will be discussed and four characteristic times will be introduced that arise due to symmetry breaking of the magnetic orientations with respect to the applied external field.

\begin{figure}
\includegraphics[width = 10cm]{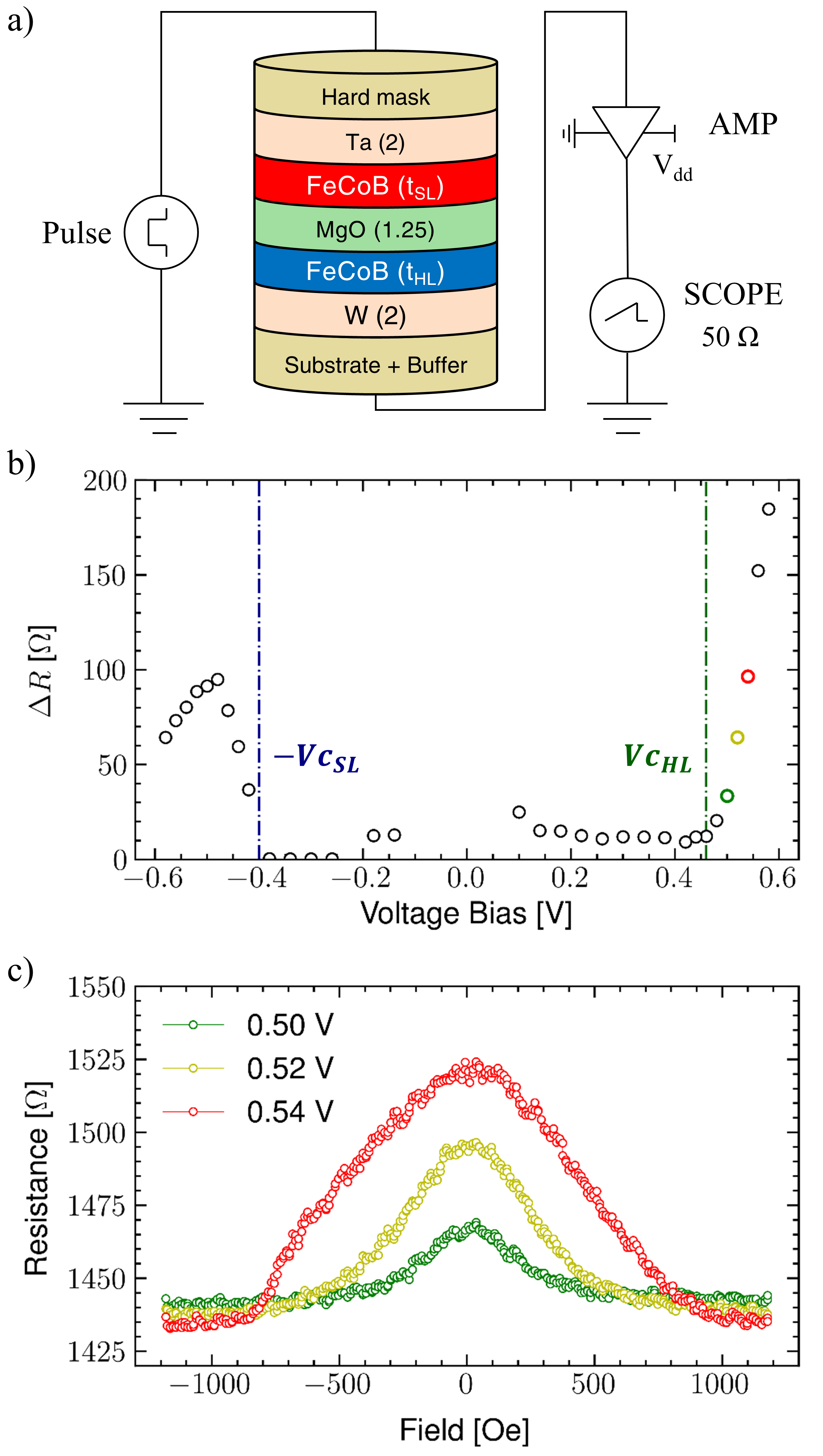}
\caption{\label{fig:2} (a) p-MTJ pillar composition where numbers inside the brackets refer to layer thicknesses in nm and schematic of the electric circuit used to sense the resistance of the device. (b) Amplitude of the resistance variation of the MTJ during a field sweep  with respect to the applied voltage. (c) Time-averaged resistance versus out-of-plane field loop at different voltage bias. Measurements performed on a pillar with a diameter of 90 nm, $t_{SL}$= 1.36 nm and $t_{HL}$= 0.81 nm.}
\end{figure}

In order to experimentally observe the spiking dynamics, p-MTJs with the composition shown in Fig.~\ref{fig:2}.a. are patterned into nanopillars of diameters ranging between 70 and 150 nm.
The stack is deposited on thermally-oxidized single crystal Si(100) by DC magnetron sputtering with a base pressure of 10$^{-8}$ mbar. On a 100mm wafer, the top (t\textsubscript{SL}) and bottom (t\textsubscript{HL}) FeCoB layers are deposited in the form of two orthogonal wedges, with thicknesses varying from 1.1 to 1.6 nm and from 0.6 to 1.0 nm, respectively. The thickness wedges allow the fabrication of devices with varying combinations of SL and HL thicknesses. Devices were patterned using Reactive Ion Etching (RIE) to create a Ta hardmask followed by an Ion Beam Etching (IBE) step to define the magnetic tunnel junction pillar. The resistance-area (RA) product is around 9 $\Omega.\mu$m² and TMR is 33\% at low bias voltage. The device used in this demonstration has an electrical diameter of 90 nm and a parallel resistance of $R_p$=1480 $\Omega$. The thinner FeCoB layer of the HL (t\textsubscript{HL} = 0.81nm) compared to the SL (t\textsubscript{SL} = 1.36nm) provides the desired asymmetry in effective anisotropy between the two layers.

Before we discuss the time resolved measurements, we present quasi-static measurements of the magneto-resistance sweeping the applied magnetic field at different voltage biases. The resistance of the MTJ is measured in real time by sensing the current through the device with an oscilloscope using the voltage drop created on its internal 50 $\Omega$ resistance after a 5$\times$ gain amplification (Fig.~\ref{fig:2}.a).
The corresponding time-averaged resistance versus field curves are shown in Fig.~\ref{fig:2}.c and TMR versus voltage curve in Fig.~\ref{fig:2}.b.
At low bias voltage, only the parallel state (P) is stabilized due to the dipolar coupling between the two layers. 
When increasing the positive bias voltage above a certain threshold ($V_{c,HL}$), the spin-polarized current provides enough torque to overcome the perpendicular anisotropy of both layers and their dipolar interaction, resulting in the dynamics described in Fig.~\ref{fig:1}. 
In this dynamical regime, the presence of high resistance spikes increases the average resistance of the device.
As shown in Fig.~\ref{fig:2}.b, the device does not experience exactly the same response depending on the bias voltage polarity. 
In the negative polarity, the electrons flow from the soft layer (SL) to the hard layer (HL). 
Since for the device presented here, the critical voltage $V_{c,SL}$ is slightly lower, with $|V_{c,SL}| < |V_{c,HL}|$ the spin current first flips the SL which results in the anti-parallel (AP) configuration. The windmill motion is then generated and stabilized when the applied voltage is larger than the critical voltage of the hard layer $V_{c,HL}$.
It is noticed that, when the device has a small asymmetry between the effective anisotropy of the two layers, the two critical voltages are very close. Upon increasing the asymmetry between the two effective anisotropies, the junctions show distinct critical voltages $V_{c,SL}$ and $V_{c,HL}$ (see supplementary material 2). 
By applying an out-of-plane magnetic field, the windmill motion progressively turns into the saturation of the P state.
The symmetry observed between positive and negative values of the applied magnetic field supports the claim that both layers are switching in the dynamical region.
In fact, if one assumes that the dynamics comes from one layer only, an external magnetic field applied along the z-axis would result in a preferential P or AP state of the MTJ, which is not the case in our device.
Additionally, we observe that the dynamics is more robust as the voltage bias increases, which is consistent with a spin-transfer torque driven reversal mechanism.

\begin{figure}
\includegraphics[width=\textwidth]{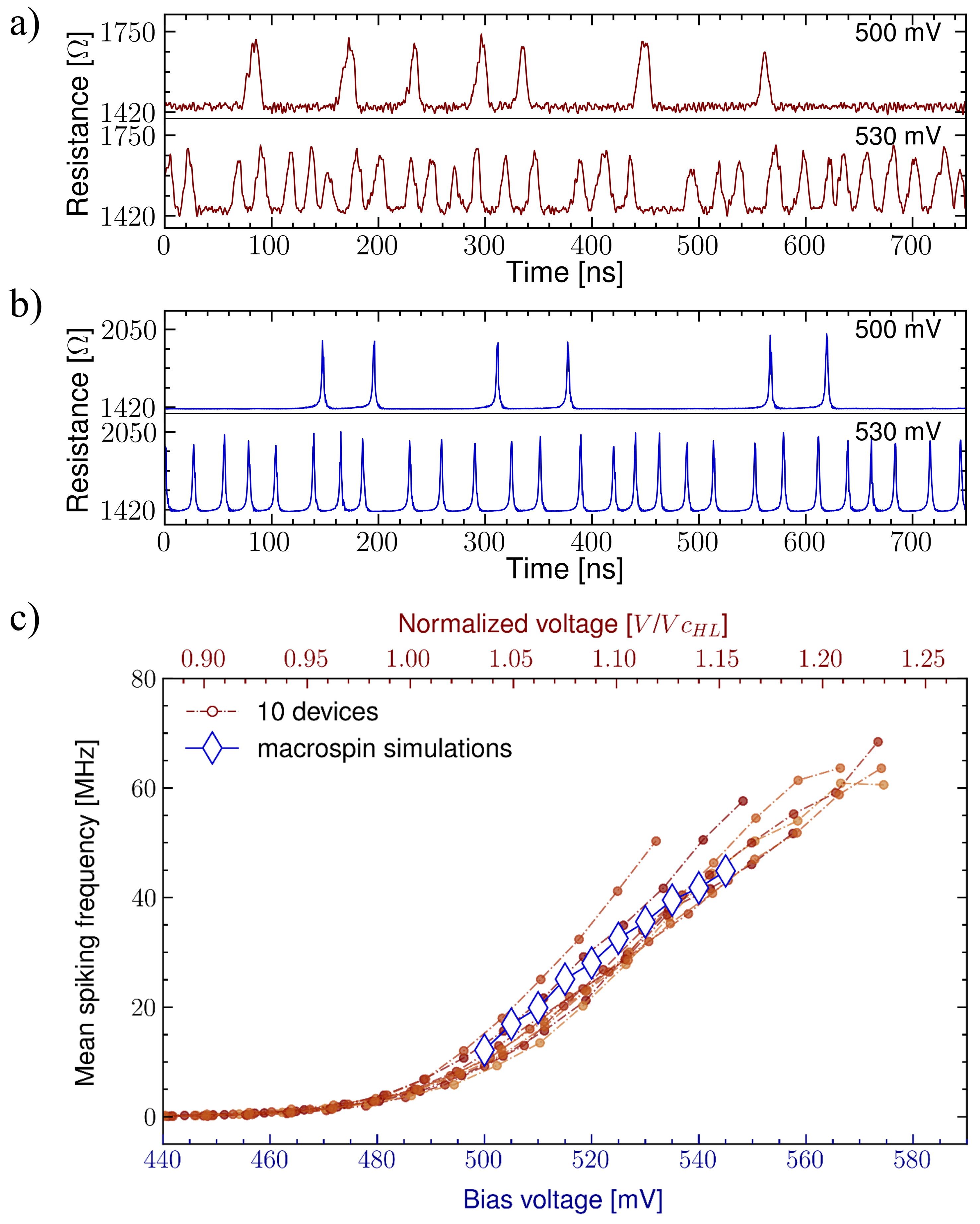}
\caption{\label{fig:3} Experimental (a) and simulated (b) time traces of a p-MTJ under a bias voltage of 500 and 530 mV. (c) Mean spiking frequency for 10 experimental devices and macrospin simulation as function of the bias voltage.}
\end{figure}

Time-resolved measurements were performed to address the dynamic features which confirm the origin of the resistance increase related to the continual switching between P and AP states. Typical time traces of a p-MTJ are shown in Fig.~\ref{fig:3}.a under a bias voltage of 500 and 530 mV. Measurements were performed on the same device as in Fig.~\ref{fig:2}.
As expected, we retrieve short AP-state lifetimes and long P-state lifetimes, characteristic of a spiking response as the one shown in Fig.~\ref{fig:1}.c. 
To retrieve the mean spiking rate of our MTJ-neuron, we binarize the signal acquired into AP ($\tau_{AP}$), P ($\tau_{P}$) states  and transitions between those states ($\tau_{Transition}$). The spiking rate is then calculated as $1 / (\tau_{AP}+\tau_{P}+2*\tau_{Transition})$.
Due to the predominant value of $\tau_{P}$, the mean spiking rate behaves similarly to the inverse of the incubation time ($f \approx 1/\tau_{P}$) and evolves non-linearly with the voltage bias (Fig.~\ref{fig:3}.c). This particular feature of the neuron activation function is a key ingredient needed in neural network application.
The time-resolved measurements show very similar behaviour for 10 devices with diameter ranging from 70 to 150 nm. 
Macrospin simulations shown in Fig.~\ref{fig:3}.b (parameters given in supplementary material 1) and previous study \cite{matsumoto_chaos_2019} also demonstrate this behaviour with applied voltage and rule out the hypothesis of a purely super-paramagnetic stochastic switching response \cite{sengupta_magnetic_2016, bapna_current_2017,parks_superparamagnetic_2018, camsari_stochastic_2017} to explain our device behaviour.

\begin{figure}
\includegraphics[width=12cm]{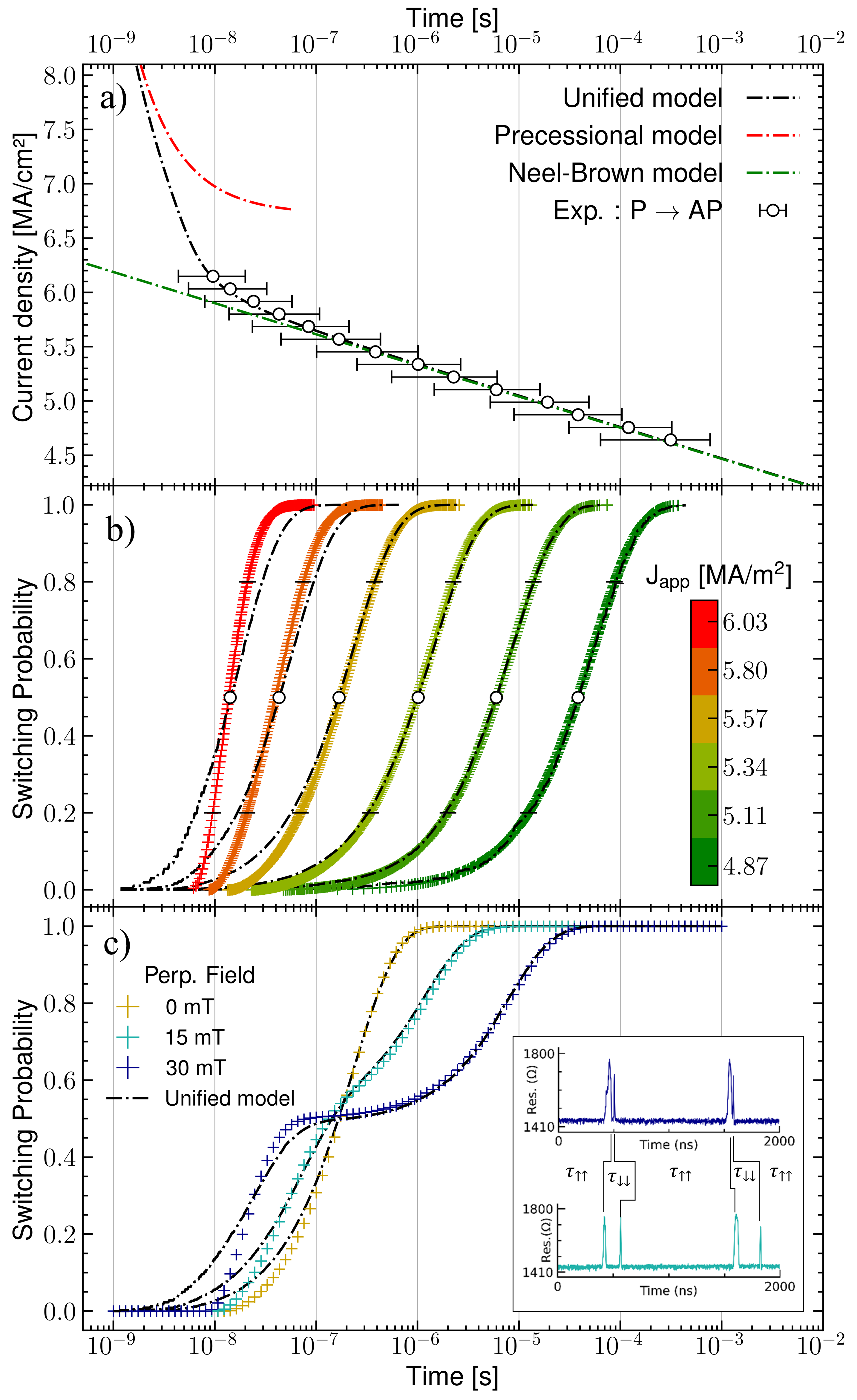}
\caption{\label{fig:4} a) Median switching time of P$\rightarrow$AP switching and b) the corresponding CDF for different injected current density. (c) CDF of the P$\rightarrow$AP switching under 5.5 $MA/m^2$ of current density with 0, 15 and 30 mT of external magnetic field. Inset of (c) shows time traces of the resistance at 15 and 30 mT of applied magnetic field.}
\end{figure}

In order to assess the capabilities of large-scale neural networks and to design appropriate hardware circuits, it is essential to have adequate compact models. For this purpose, we have developed a compact model that depends on the critical parameters of our MTJ device.
As the output mean frequency depends mainly on the incubation time, the model specifically focuses on how the incubation time changes with voltage bias and perpendicular external magnetic field.
The incubation time between spikes varies significantly in the low current regime due to the stochastic nature of the STT switching in that regime. Below the critical bias voltage ($V_{c,HL}$), the dependence of this parameter with the applied current density is exponential \cite{brown_thermal_1963}. Above $V_{c,HL}$, the incubation time falls into a precessional regime \cite{sun_spin-current_2000,koch_time-resolved_2004, he_switching_2007} with a 1/V dependency.
To combine the two regimes, we extended the model proposed by Yang \textit{et al.} \cite{yang_universal_2022} by adding the term $\frac{H_{dip}+H_{app}}{H_{K}}$ in the equation of the switching time to consider also the impact of an external magnetic field, as it has been proposed previously \cite{tomita_unified_2013}.
The incubation time $\tau_P$, related to the switching of the hard layer (HL), is expressed as function of the current density $J$ and of the perpendicularly applied field $H_{app}$ for a given switching probability $P$ :
\begin{eqnarray}
\tau_P(J, H_{app}) = \dfrac{\boldsymbol{A}_P \cdot \exp\Big[(1-k)\cdot\Delta\cdot j\cdot h^{2} \Big]+k\cdot\Delta\cdot\boldsymbol{B}_P}{\ln\Big[1+\exp\Big(-k\cdot\Delta \cdot j \cdot h^{2}\Big)\Big]},
\label{eq:one}
\end{eqnarray}
where : $\boldsymbol{A}_P = -\tau_{0}\cdot\ln(1-P) $, $ \boldsymbol{B}_P = \frac{1}{2}\cdot\tau_{D}\cdot\ln\big(-\frac{4\Delta}{\ln P}\big) $, $j = 1- \frac{J}{J_{c0}}$ and $ h = 1-\frac{H_{dip}+H_{app}}{H_{K}}$. \\
The $\boldsymbol{A}_P$ term describes the switching time in the low current density regime (Neel-Brown model). The second term $\boldsymbol{B}_P$ is representative of the precessional mode achieved in the higher current range. The spin-torque intensity $k$, proposed by Yang \textit{et.al}\cite{yang_universal_2022}, is used to merge the two models in a single expression and varies linearly from 0.25 to 0.15 in the range of the current density applied. The energy barrier $\Delta$ (46 $k_BT$), the characteristic times, $\tau_0$ (100 ns) and $\tau_D$ (270 ns), and the critical current density $J_{c0}$ (6.71 $MA/m^2$) are extracted from the fit of the cumulative distribution function (CDF) with the unified model Eq.~(\ref{eq:one}) for current densities between 4.8 and 6.2 MA/cm². Fig.~\ref{fig:4}.a shows the median of the switching time from P state to AP state and the error bars correspond to the 20-80\% range of the switching probability of the CDFs in Fig.~\ref{fig:4}.b. In the thermally assisted reversal case with long incubation times, the model accurately describes the experimental data. On the contrary, for very short lifetimes ($ J > 5.8 MA/m^2$), the fitting of the full CDF starts to deviate from device data. This can be explained to some extent by the 350 MHz bandwidth limitation of the amplifier used to provide signal gain, limiting the accuracy of measurements closer to the precessional regime.

As mentioned above, our dual structure is subject to the dipolar coupling. To evaluate its role, we performed systematic experiments under an external magnetic field to extract the dipolar field $H_{DIP}$ (12.2 mT) and the anisotropy field of the hard layer $H_K$ (147 mT).
The external field affects the shape of the CDF since it brakes the symmetry between the two P states ($\downarrow\downarrow$, $\uparrow\uparrow$) and the two AP states ($\downarrow\uparrow$, $\uparrow\downarrow$), leading to four different characteristic times : $\tau_{\downarrow \downarrow}, \tau_{\downarrow \uparrow}, \tau_{\uparrow \uparrow}, \tau_{\uparrow \downarrow}$. At a given current density and for a positive applied magnetic field (+z), $\tau_{\downarrow \downarrow}$ and $\tau_{\downarrow \uparrow}$ times decrease because the field polarity helps magnetization reversals from $\downarrow$ to $\uparrow$ configuration due to the Zeeman energy. Whereas the $\tau_{\uparrow \uparrow}$ and $\tau_{\uparrow \downarrow}$ times increase because the positive applied field hinders the magnetization to switch from $\uparrow$ to $\downarrow$ orientation. 
The consequences of the external applied field on the signal acquired are shown in the inset of Fig.~\ref{fig:4}.c where a long P state ($\uparrow \uparrow$) is followed by a shorter P state ($\downarrow \downarrow$). The asymmetries between the two P states and the two AP states increase with the field amplitude. The plateau at 50\% of probability in Fig.~\ref{fig:4}.c confirms that there is the same number of short and long P states when the applied field significantly changes the associated characteristic times $\tau$. This is an additional proof that the windmill dynamics drives the spiking response of the device. 

The robustness of the windmill dynamics with respect to the external magnetic field is an important requirement for device integration as both their number and density increase. Indeed, neuromorphic computing generally requires millions of neurons and synapses in a limited area for embedded applications which may create stray field interactions between devices. 
Even if the shape of the CDF is modified with external applied magnetic field, the spiking feature remains, as confirmed by the time traces in the inset of Fig.~\ref{fig:4}.c. From another viewpoint, one may see the external field as an additional degree of freedom to tune the probability distribution of the spiking signal. 
To conserve the high energy efficiency \cite{grollier_neuromorphic_2020} of analog neuromorphic circuits with respect to other computing schemes, the energy consumption at the device level must be evaluated. The energy needed to produce one single spike in our device, was calculated as $E = 1/f*V_{bias}^2/R $, showing values of 4 to 16 pJ in a frequency range from 10 to 50 MHz. State-of-art STT-RAM achieves nowadays switching reversal below 100 fJ \cite{wang_low-power_2013, zeng_effect_2011} with high stability ($\Delta > 60$). This energy per operation could further decrease by reducing the effective magnetic anisotropy of the layers as our proposed device does not require any high stability layer.
Furthermore, several issues arising from the SAF element in conventional STT-RAM are avoided, such as the use of critical materials \cite{palomino_evaluating_2021}, the additional complexity to combine different crystal orientations \cite{djayaprawira_230_2005} and the limited thermal budget due to atom migration \cite{karthik_transmission_2012}.
A CMOS-compatible two-terminal device with a simple material stack, not requiring a SAF reference, and operating without set/reset or compensation applied magnetic fields are strong benefits of our solution. These demonstration of the spiking functionality with device diameters down to 70 nm opens a concrete path to design dense arrays for in-memory computing.

In summary, we demonstrated the use of a dual free layer MTJ to generate electrical spiking signals under an applied bias voltage. The output spiking frequency is a non-linear activation function with respect to the external excitation as required in NN algorithms. Simulations and multiple experimental data sets converge to explain the origin of the spikes as the generation of a windmill-like magnetization reversal sequence.
Towards hardware circuit design, we develop a compact model that describe our experimental data combining both thermally activated and precessional switching regimes. The possible impact of stray fields in dense arrays was shown to be negligible, as demonstrated by the resilience of the windmill dynamic under applied magnetic field up to 30 mT. The low power consumption (4-16 pJ/spike) and high scalability of this MTJ-based spiking neuron ($<$ 80 nm) presents a concrete solution for the implementation of SNN in hardware at the nanoscale.

\begin{acknowledgement}
This work was supported by ANR via grant SpinSpike Projet-ANR-20-CE24-0002.
\end{acknowledgement}

\begin{suppinfo}
The data that support the findings of this study are available from the corresponding author upon reasonable request.
\end{suppinfo}

\bibliography{Paper_letter}

\end{document}